\documentclass[fleqn,usenatbib]{mnras}

\usepackage[T1]{fontenc}
\usepackage[normalem]{ulem} 
\DeclareRobustCommand{\VAN}[3]{#2}
\let\VANthebibliography\thebibliography
\def\thebibliography{\DeclareRobustCommand{\VAN}[3]{##3}\VANthebibliography}

\newcommand{\msun}{\mbox{M$_{\odot}$}}
\newcommand{\fss}{\hbox{$.\!\!^\mathrm{s}$}}        
\newcommand{\h}{$^\mathrm{h}$}
\newcommand{\m}{$^\mathrm{m}$}

\newcommand{\nh}{$N_\mathrm{H}$}

\newcommand{\psr}{J1957}
\newcommand{\jpsr}{J1957+5033}
\newcommand{\ergs}{erg~s$^{-1}$}

\newcommand{\phs}{ph~cm$^{-2}$~s$^{-1}$~keV$^{-1}$}
\newcommand{\xmm}{\textit{XMM-Newton}}
\newcommand{\chan}{\textit{Chandra}}
\newcommand{\fermi}{\textit{Fermi}}
\newcommand{\degs}{\ifmmode ^{\circ}\else$^{\circ}$\fi}
\newcommand{\amin}{\ifmmode ^{\prime}\else$^{\prime}$\fi}
\newcommand{\tc}{t_\mathrm{c}}

\usepackage{graphicx}	
\usepackage{amsmath}	
\usepackage{amssymb}	
\usepackage{threeparttable}

\title[PSR J1957+5033 in the optical]{
Likely optical counterpart of the cool middle-aged 
pulsar J1957+5033
}

\author[D. A. Zyuzin et al.]{D. A. Zyuzin$^{1}$\thanks{E-mail:da.zyuzin@gmail.com},
S. V. Zharikov$^2$,
A. V. Karpova$^1$,
A. Yu. Kirichenko$^{2,1}$,
Yu. A. Shibanov$^1$,
S. Geier$^{3,4}$,\newauthor
A. Yu. Potekhin$^1$,
V. F.  Suleimanov$^{5,6}$
and A. Cabrera-Lavers$^{3,4}$\\
$^1$Ioffe Institute, Politekhnicheskaya 26, St. Petersburg, 194021,  Russia \\
$^2$Universidad Nacional Aut\'{o}noma de M\'{e}xico, Instituto de Astronom\'{i}a, AP 106,  Ensenada 22800, BC, M\'{e}xico \\
$^3$Instituto de Astrof\'isica de Canarias, V\'ia L\'actea s/n, E38200, La Laguna, Tenerife, Spain\\
$^4$GRANTECAN, Cuesta de San Jos\'e s/n, E-38712, Bre\~{n}a Baja, La Palma, Spain\\
$^5$Institut f\"{u}r Astronomie und Astrophysik, Sand 1, D-72076 T\"{u}bingen, Germany\\
$^6$Kazan (Volga region) Federal University, Kremlevskaja str., 18, Kazan 420008, Russia \\
}

\date{Accepted XXX. Received YYY; in original form ZZZ}

\pubyear{2021}

\begin{document}
\label{firstpage}
\pagerange{\pageref{firstpage}--\pageref{lastpage}}
\maketitle

\begin{abstract}
The 840 kyr old  
pulsar PSR
\jpsr, detected so far only in $\gamma$- and X-rays,   
is  a  nearby
and rather cool neutron star with a temperature of 0.2--0.3 MK, a distance  of $\la 1$~kpc, 
and a small colour reddening excess $E(B-V) \approx 0.03$. 
These properties make it an ideal candidate to detect in the optical
to get additional 
constraints on its parameters.
We thus performed the first  deep
optical observations of the pulsar with the 10.4-meter Gran Telescopio Canarias 
in the $g'$ band and found its possible  counterpart with
$g'=27.63\pm 0.26$. The counterpart candidate position is consistent with the  X-ray coordinates of the pulsar
within the 0.5 arcsec accuracy. 
Assuming that this is the real counterpart, we analysed the pulsar X-ray spectrum 
together with the derived optical flux density. 
As a result, we found that the 
thermal emission from the bulk surface of the cooling neutron star
can significantly contribute to its optical flux. 
Our multi-wavelength spectral analysis favours the pulsar nature of the detected optical source, since it provides physically adequate  parameters of the pulsar emission. 
We show that the optical data can provide new constraints
on the pulsar temperature and distance.
\end{abstract}




\section{Introduction}
\label{sec:intro}

The radio-quiet $\gamma$-ray pulsar PSR \jpsr\ (hereafter \psr) belongs to the family of the so-called middle-aged
neutron stars (NSs) with characteristic ages between $\sim 10^4$ and $\sim 10^6$ yr. Their spectra in soft X-rays 
and ultraviolet (UV) are typically dominated by thermal emission components from  bulk surfaces of  cooling NSs. 

\psr\ was 
discovered with the \fermi\ Gamma-ray Space Telescope \citep{SazParkinson2010}. 
It has a  spin  period\footnote{The spin period 
and the period derivative are taken from the \psr\ timing solution 
which can be found at 
\url{https://confluence.slac.stanford.edu/display/GLAMCOG/LAT+Gamma-ray+Pulsar+Timing+Models}. 
See also \citet{kerr2015} for details.}  $P=375$ ms, 
a period derivative $\dot{P}=7.1\times10^{-15}$ s~s$^{-1}$, 
a characteristic age $\tc\equiv P/2\dot{P}\approx 840$ kyr, 
a spin-down luminosity $\dot{E}=5.3\times10^{33}$ \ergs,
and a spin-down magnetic field $B=1.65\times10^{12}$ G.
Based on the 25-ks \chan\ data set of the pulsar field, \citet{marelli2015} proposed the pulsar 
X-ray counterpart and showed that its spectrum 
in the 0.3 -- 10 keV band can be described by an
absorbed  
power law (PL) model with a
photon  index $\Gamma \sim 2.1$.
\citet{XMMPSRJ1957} performed 87-ks \xmm\ observations of the X-ray source and confirmed its pulsar nature
by detection of X-ray pulsations with the 
 spin period of the NS. They also found a faint trail-like spatial feature
which  extends from the pulsar towards North-West for $\sim$8 arcmin.
It is likely associated with the  
pulsar wind nebula (PWN) or a misaligned outflow of relativistic particles generated by the pulsar. 
Furthermore, the data analysis revealed that an additional 
soft thermal component is required to describe the pulsar spectrum in the 0.15--0.5 keV range. 
It can originate from the entire surface of the 
NS with the dipole magnetic field and respective non-uniform temperature distribution covered by 
a
hydrogen atmosphere. Using the atmosphere model in the spectral fit, the derived  
effective temperature of  \psr\ is $0.2-0.3$ MK, 
making this pulsar the coldest among middle-aged NSs 
with measured surface temperatures \citep{potekhin2020}.
Alternatively, the blackbody (BB) model with a larger temperature of $\approx0.6$ MK can also fit the thermal spectral component well, implying that it originates from some part of the NS surface \citep{XMMPSRJ1957}. This may be reminiscent of the emission from a solid state surface of the NS. 

\begin{figure*}
\setlength{\unitlength}{1mm}
\resizebox{15.cm}{!}{
\begin{picture}(130,72)(0,0)
\put (-15,0){\includegraphics[width=7.5cm, bb = 0 0 750 750, clip=]{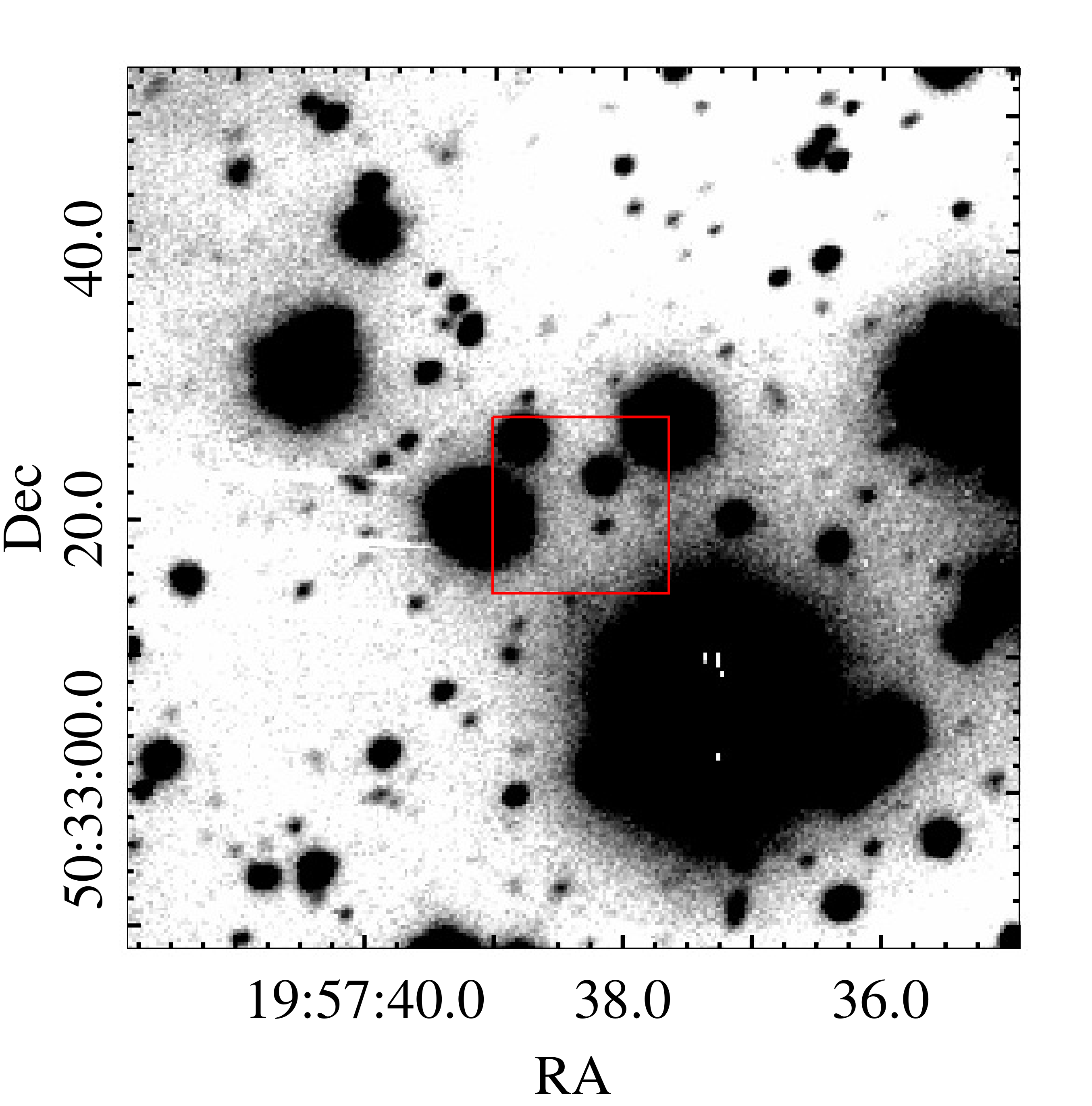}}
\put (65,5){\includegraphics[width=7.8cm, clip=]{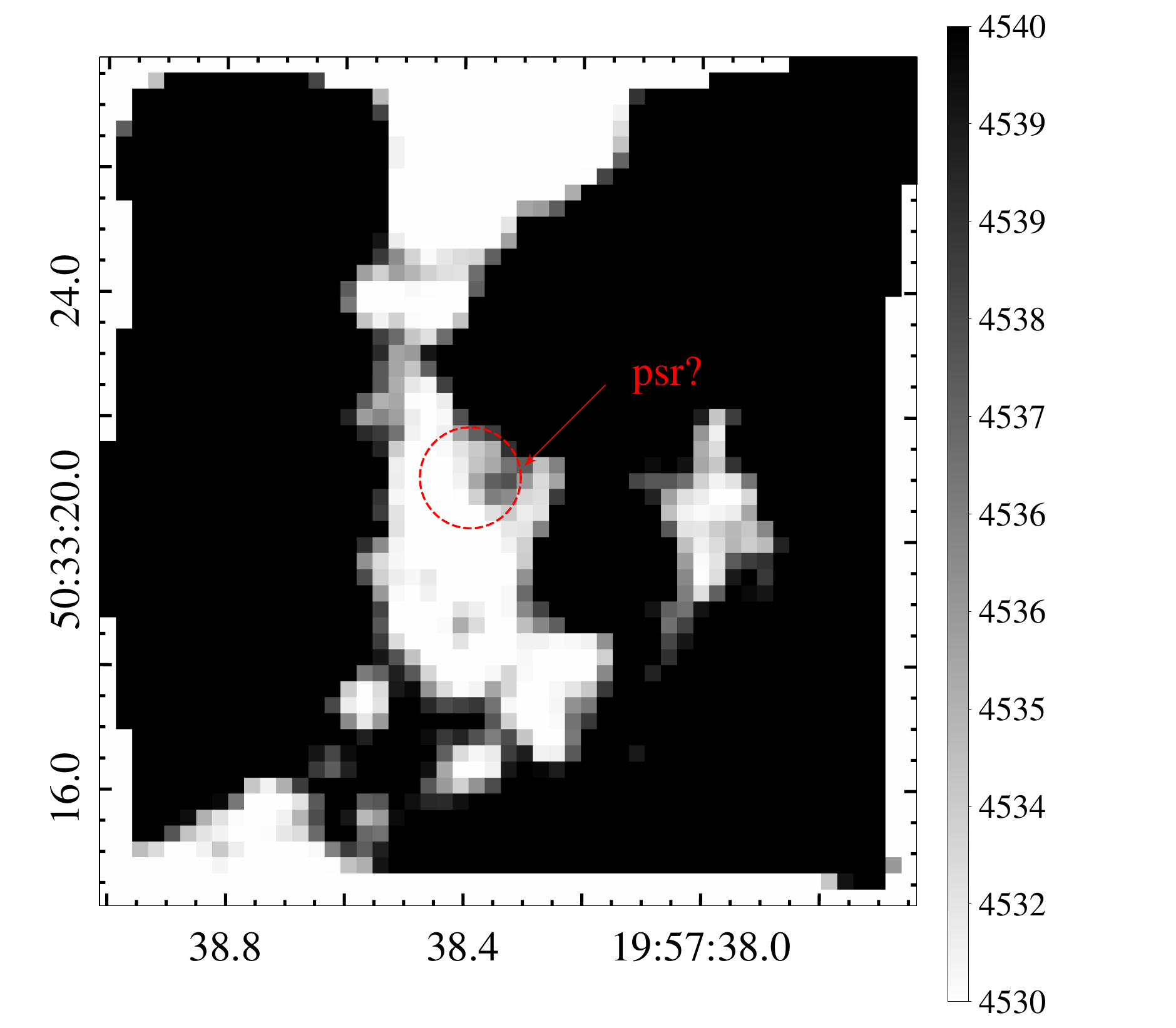}}
\end{picture}}

\caption{The left panel shows the fragment of the 
stacked
GTC/OSIRIS Sloan $g'$-band image around the pulsar position. The region within the  box with the size of $\sim$13 arcsec $\times$ 13 arcsec containing the pulsar is enlarged in the  right panel where  
the circle with the 0.81 
arcsec radius represents the 90 per cent 
uncertainty of the pulsar X-ray position.  
There is a faint object within it marked by the arrow which is proposed as the pulsar optical counterpart. The right image is smoothed with a two pixel Gaussian kernel. The  vertical
colour-bar represents the mapping of data 
values in counts to colours. The seeing  is 1.1 arcsec.
}
\label{fig:psr}
\end{figure*}
The distance to the pulsar estimated using the interstellar 
extinction--distance relation in X-ray spectral fits as a prior remains rather uncertain, 0.1--1 kpc \citep{XMMPSRJ1957}. 
The small distance and the long wavelength extrapolation of the  thermal emission X-ray spectral component of \psr\ assuming the atmosphere model fit  
show that it can be sufficiently bright to be visible in the optical and UV. This is also supported by the 
low X-ray absorption column density \nh\ towards the pulsar of about 3$\times$10$^{20}$ cm$^{-2}$ implying the colour excess $E(B-V) \approx 0.03$ in accord to the relation by \citet{foight2016}. 
Optical-UV observations 
could  potentially test the models of 
the thermal emission of \psr. 
They
could be also important to better 
constrain the temperature, parameters of the PL emission component,
the  distance  and luminosity of \psr. Such work was done only for a very limited number of NSs showing that the optical-UV data can be indeed very useful in this respect \citep[e.g.][]{RXJ1856.5-37542007,Zharikov2021}. 
The deepest available optical data on the \psr\ field were obtained by Pan-STARRS \citep{2016arXiv161205560C} and  \citet{beronya2015}.  They do not reveal any pulsar counterpart down to the 23 magnitude upper limit.

Therefore, we performed a 
deeper optical  search of  
the pulsar counterpart using  the 10.4-meter Gran Telescopio Canarias (GTC). 
In this paper we present results of these observations.
The data and their reduction are described in Section~\ref{sec:data} and results -- in Section~\ref{sec:results}.
The multi-wavelength spectral analysis is presented in Section~\ref{sec:mw}.
The results are discussed and summarised in Section~\ref{sec:discussion}.


\section{Observations, data reduction, and calibration} 
\label{sec:data}

The \psr\ field was observed\footnote{Programme GTC13-20BMEX, PI S. Zharikov} on 
13 September 2020 during $\sim2$ h under clear and dark sky using 
the Sloan $g'$ filter of the Optical System for Imaging and 
low-intermediate Resolution Integrated Spectroscopy 
(OSIRIS\footnote{\url{http://www.gtc.iac.es/instruments/osiris/}})  
instrument at the GTC. OSIRIS consists of two CCDs with a field of view (FoV) of 7.8 arcmin~$\times$~7.8 arcmin  and a pixel scale of 0.254 arcsec (with binning 2x2).
The pulsar was exposed on CCD2 with a mean airmass of 1.17. 
To avoid being affected
by bad pixels, we used five arc-second dithering
between 200~s individual exposures. 
In addition, three short 15--20 s exposures of the pulsar field in the $g'$ band 
were obtained to avoid saturation of bright stars that were further 
used for precise astrometry. During the observations, the seeing varied between 0.96 and 1.36 arcsec.

We performed standard data reduction, including bias subtraction 
and flat-fielding, using the Image Reduction 
and Analysis Facility ({\sc iraf}) package \citep{iraf}. 
The cosmic rays were removed from the images with the L.A.Cosmic algorithm \citep{vandokkum}. Inspection 
of the target vicinity in each individual exposure revealed the presence of ghosts caused by nearby bright stars that fall onto 
the target position in four images. These  images were excluded from the analysis.
The rest of the 33 exposures were then aligned 
and combined resulting in a final image with the total exposure time of 6600 s and 
seeing of 1.1 arcsec.  

The photometric calibration was obtained using several stars in the CCD2 FoV from the Pan-STARRS catalogue \citep{2016arXiv161205560C}. 
To determine the magnitude zero-point, we used their measured and catalogue magnitudes and the mean OSIRIS atmosphere extinction  coefficient 
$k_{g'}$~= 0.15(2)\footnote{\url{http://www.gtc.iac.es/instruments/osiris/media/CUPS\_BBpaper.pdf}}. 
The resulting zero-point is $Z_{g'}=28.54(3)$. The average difference between the Pan-STARRS $g$ and Sloan $g'$ magnitudes is only 0.014(12) \citep{2012ApJ...750...99T}, which is within the zero-point error budget. 

The astrometric calibration  was computed using the image combined from  the short exposures and a set of 15 relatively bright  non-saturated stars with negligible proper motions from the \textit{Gaia} DR2 catalogue 
\citep{gaia1, gaia2} located in the pulsar vicinity in the GTC image.  
Their position uncertainties on the image and in the catalogue were $\lesssim 50$ mas and $\lesssim 1$ mas, respectively.   
For the astrometric referencing, we used the {\sc ccmap} routine of {\sc iraf}, which takes into account the frameshift, rotation, and scale factor. 
Formal $rms$ uncertainties of the resulting astrometric fit 
were 40 mas for both coordinates.  The resulting solution was applied to the final image. 
Allowing for  the fit and reference star position  uncertainties,    
the resulting astrometric accuracy of the final image was estimated to be 60 mas.

\section{Results} 
\label{sec:results}

In Figure~\ref{fig:psr} we present a fragment of the $g'$-band image of the   
J1957
field. A magnification of the pulsar vicinity is shown in the right panel where  
the \textit{Chandra} X-ray position of the pulsar,
R.A.~=~19\h57\m38\fss390
and Dec.~=~+50\degs33\amin21\farcs02 \citep{XMMPSRJ1957},
is shown   
by the  circle 
with a radius of 
0.81 arcsec. 
The latter includes the 
90\% 
\textit{Chandra} nominal astrometric accuracy\footnote{\url{https://cxc.harvard.edu/cal/ASPECT/celmon/}}  of  
0.8 arcsec and the optical referencing uncertainty.
There is a faint optical source  within the circle. 
It is detected with a signal 
to noise ratio of about four (see below).    
Its coordinates 
are R.A.~=~19\h57\m38\fss35(1),
Dec.~=~+50\degs33\amin20\farcs96(12), 
where numbers in brackets correspond to 1$\sigma$ uncertainties.  
We carefully inspected individual images and found that this is a real source  and not an artifact caused by cosmic rays or CCD defects. There are no CCD defects or cosmic ray tracks found. Moreover, the object is always detected even if any individual image is excluded.
Therefore, by the position coincidence, we further consider it as a likely optical counterpart of the pulsar. 
Another nearest optical source is located at $\approx$ 2.2 arcsec South-West of the  pulsar. This one and other more distant objects can hardly be  potential counterparts as they are outside of the 99 percent 
uncertainty circle of the \textit{Chandra}  pulsar position with the radius of 1.4 arcsec.       

The background around the counterpart candidate is non-uniform due to contamination from nearby bright stars. Using standard aperture photometry with a circular annulus region for background estimates is problematic in this case. Point spread function photometry is also unreliable  due to the saturation of these stars.      
To measure the source magnitude, we, therefore, applied the method described in \citet{2013MNRAS.435.2227Z}. To measure the  
source$+$background flux in counts, 
we used a square aperture of 1 arcsec $\times$ 1 arcsec centred at the   
brightest  pixel of the source. 
For the mean background and its standard deviation estimates, we used the same aperture centred at four different random positions in the nearby vicinity of  the source less affected by neighbour stars (south 
of the
pulsar position).
The aperture correction derived using several bright unsaturated stars in the image was then applied to the source count rate, which was 
finally transformed to the magnitude using the photometric zero-point and atmosphere extinction coefficient. 
As a result, 
we measured the likely 
counterpart magnitude to be $g'=27.63\pm 0.26$, corresponding  
to a flux
density of 
$f_{\rm obj}=34^{+10}_{-7}$ nJy and showing that the source is detected at  
the 4$\sigma$ level.  

If this source is an unrelated background object or some extreme fluctuation of background,  we can set a 3$\sigma$ upper limit on the pulsar brightness in the optical based on the background level and its variation near the pulsar position,  
$g'_{3\sigma} = 27.94$, 
resulting in  the flux density $f_{3\sigma} = 25$ nJy. We see that the derived counterpart candidate brightness is close to the detection limit of the observations. 

Finally, we do not detect in the optical 
any signature of the pulsar trail-like feature seen by \citet{XMMPSRJ1957} in X-rays.

\section{Multi-wavelength spectral analysis} 
\label{sec:mw}

It is instructive to compare the flux density of the likely
optical counterpart and its upper limit with available X-ray data on the pulsar.

\subsection{X-ray data}

We used the same data as in \citet{XMMPSRJ1957}, i.e. the \chan\ Advanced CCD Imaging Spectrometer (ACIS)\footnote{ObsID 14828, PI M. Marelli, observation date 2014-02-01, 25 ks exposure}    
and \xmm\ European Photon Imaging Camera 
(EPIC)\footnote{ObsID 0844930101, PI D. Zyuzin, observation date 2019-10-05} data sets.
To analyse the data,
we utilised the \xmm\ Science Analysis Software ({\sc xmm-sas})
v. 17.0.0 
\citep{sas2014ascl.soft04004S}
and \chan\ Interactive Analysis of Observations ({\sc ciao})
v. 4.12 \citep{ciao}   packages.
The \xmm\ data suffer from background flaring, and after cleaning
the effective exposures are 79.8, 79.8 and 48.7 ks for EPIC-MOS1, -MOS2 and -pn
detectors, respectively. 
The pulsar spectrum was extracted 
from \xmm\ and \chan\ data using
{\sc evselect} and 
{\sc specextract} tools, respectively.
The resulting number of source counts after background 
subtraction is 
232(MOS1) + 254(MOS2) + 902(pn) + 88(ACIS).
 The full description of data reduction and extraction of the spectra can be found in \citet{XMMPSRJ1957}.

\begin{figure}
\begin{minipage}[h]{1\linewidth}
\center{
\includegraphics[width=1\linewidth,clip]{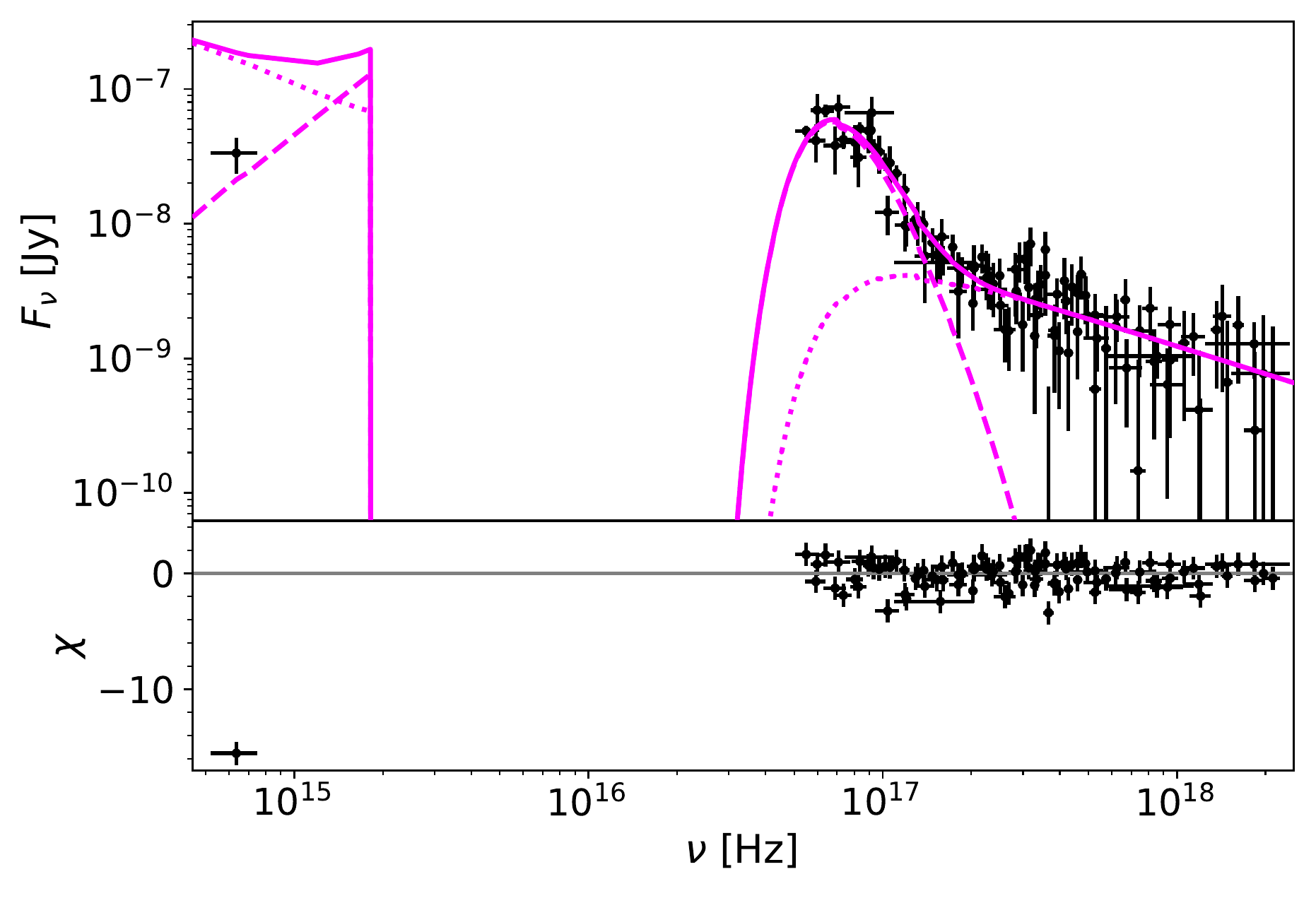} 
\includegraphics[width=1\linewidth,clip]{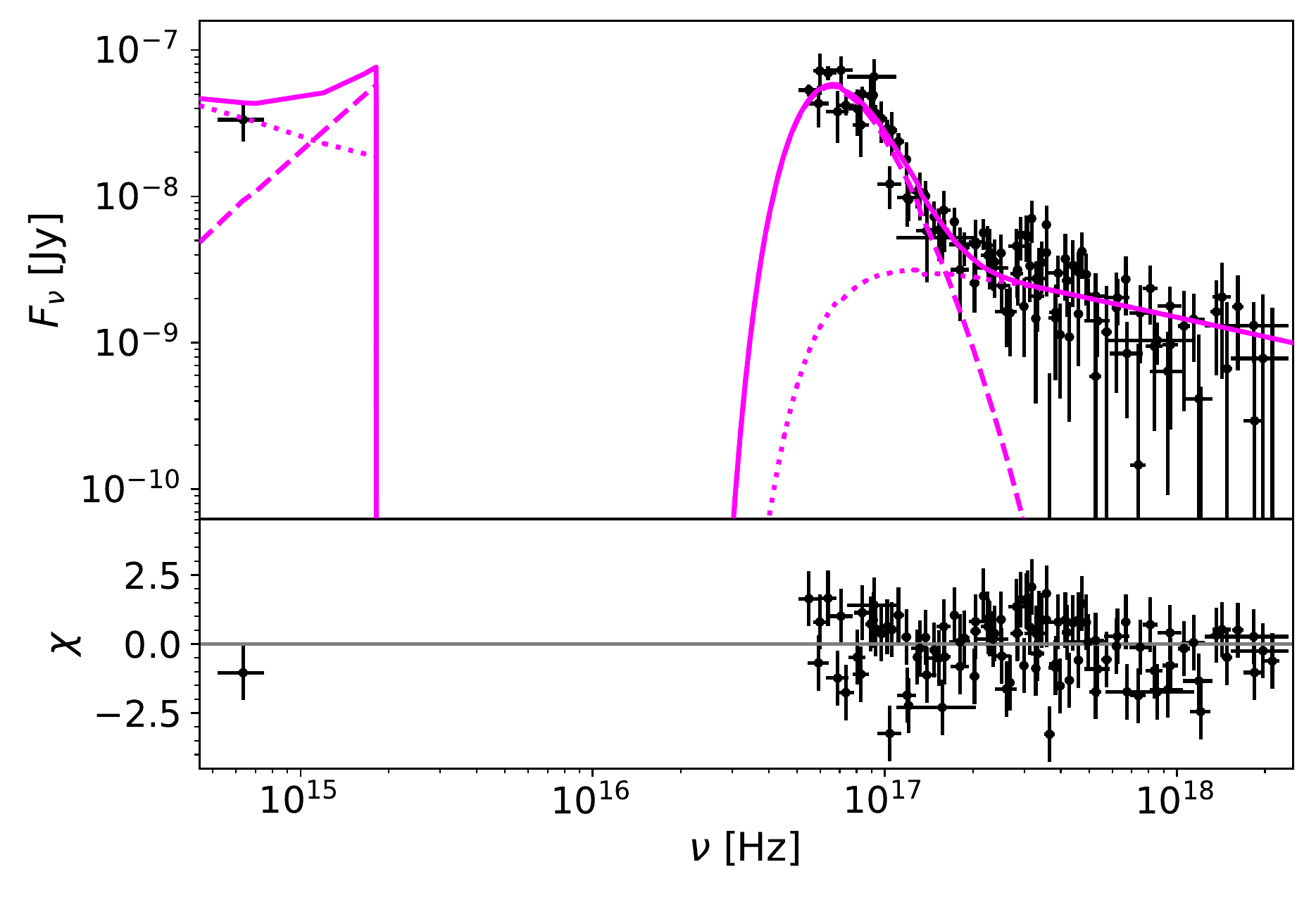} 
\includegraphics[width=1\linewidth,clip]{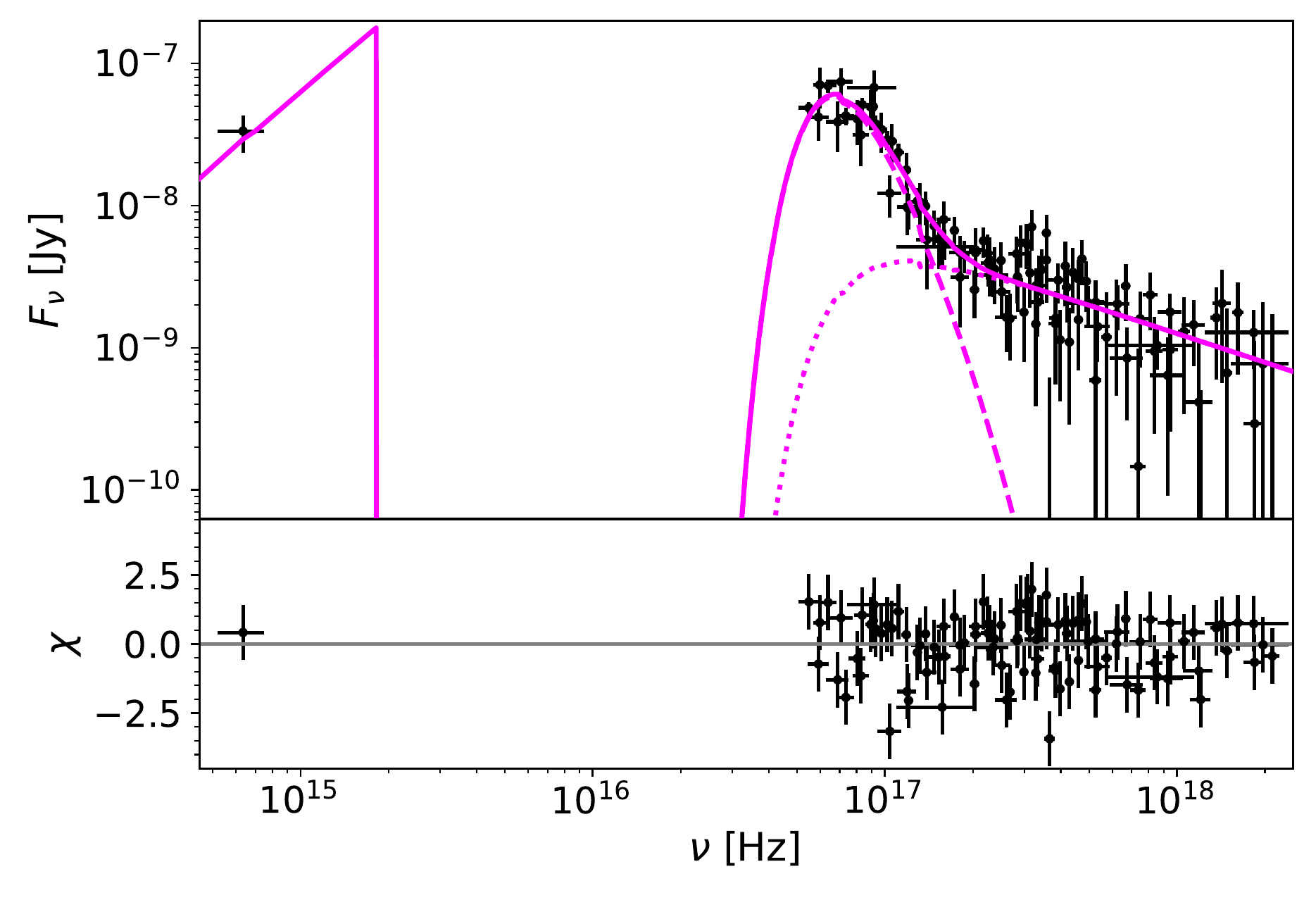} 
}
\end{minipage}
\caption{ Observed spectra of \psr\ in the optical--X-ray range obtained with \chan, \xmm\ and GTC, best-fitting model {\sc nsmdip1}+PL and
 the fit residuals. 
The model is shown by the solid lines while its thermal and non-thermal components -- 
by the dashed and dotted lines, respectively. 
In the top panel the best-fitting model of the X-ray spectra alone is 
simply extrapolated to the optical. 
In the middle panel the optical--X-ray data are fitted simultaneously
assuming common PL parameters (case 1) while in the bottom panel the PL normalization
is set to zero in the optical (case 2). For illustrative purposes, 
the X-ray spectra were grouped to ensure at least 10 counts per energy bin. }
\label{fig:psr-models-src}
\end{figure}

\begin{figure*}
\begin{minipage}[h]{1\linewidth}
\center{
\includegraphics[width=0.9\linewidth,clip]{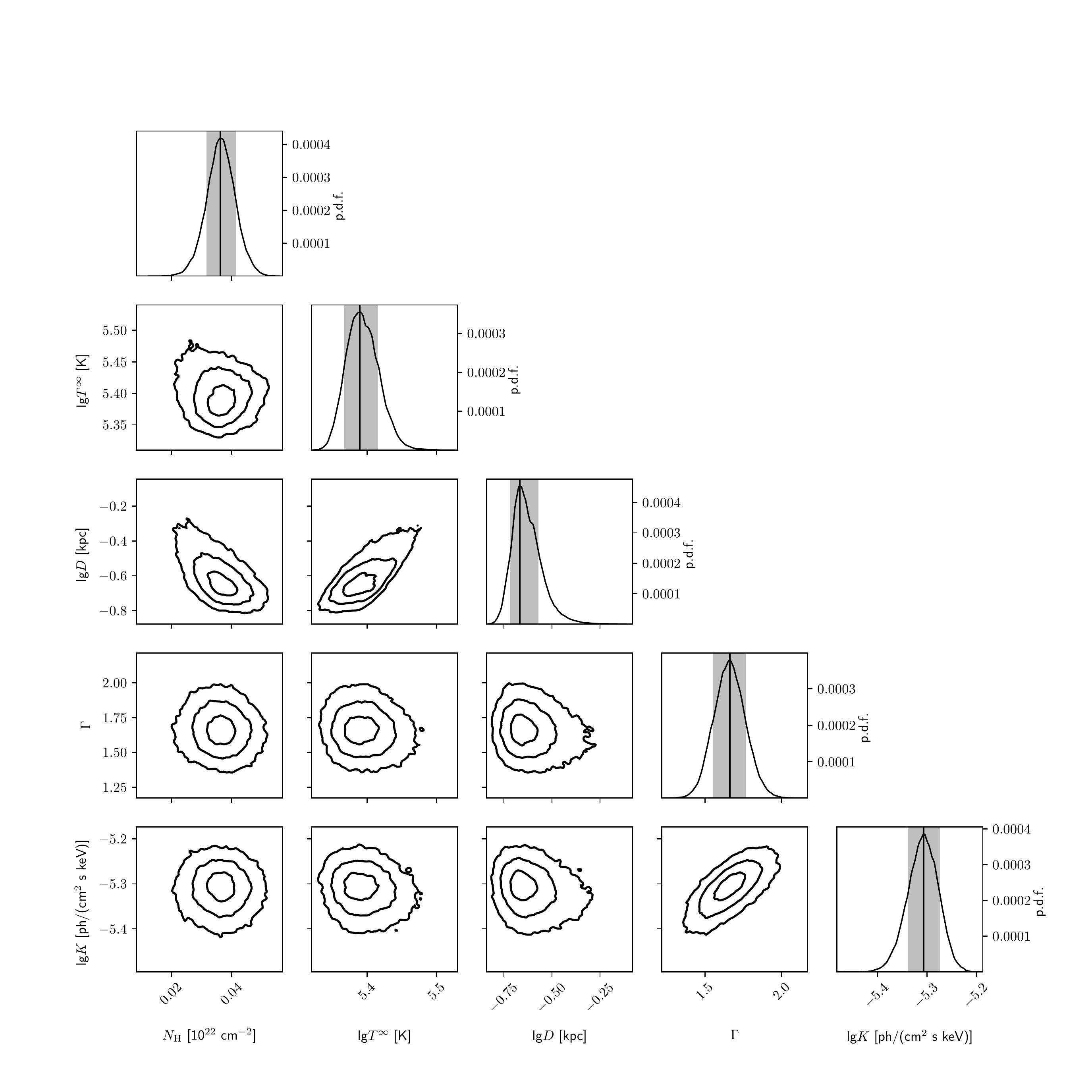} 
}
\end{minipage}
\caption{1D and 2D marginal posterior distribution functions for parameters in case 2 
(see Table~\ref{tab:fit}). In 1D distributions, the solid vertical lines and the grey strips
show the best-fitting values and 68 per cent credible intervals. In 2D distributions, 
39, 86 and 99 per cent confidence contours are presented. }
\label{fig:trngl}
\end{figure*}

\subsection{Spectral analysis} 
\label{sec:spec}

For the spectral analysis we used the X-Ray Spectral Fitting Package 
({\sc xspec}) v.12.10.1 \citep{arnaud1996}. 
To consider the optical data-point together with the X-ray spectra, 
the measured optical flux density $f_{\rm obj}$ was used to generate 
respective count rate 
and response files 
by applying the {\sc ftools}\footnote{\url{http://heasarc.gsfc.nasa.gov/ftools}; 
\citep{ftools}} task {\sc ftflx2xsp}.
The extracted X-ray spectra were fitted in 0.3--10 keV, 0.2--10 keV and 0.15--10 keV ranges for the \chan, 
MOS and pn data, respectively. 

To take into account the absorption in the interstellar medium (ISM), 
we used the Tuebingen-Boulder model {\sc tbabs} 
with the {\sc wilm} abundances \citep*{wilms2000} for the X-ray spectra 
while for the optical point the {\sc redden} model was applied.
The latter uses the extinctions from \citet*{cardelli1989}. 
The ISM optical colour reddening excess $E(B-V)$ was linked with the equivalent hydrogen absorbing column density
\nh\ in X-rays by the empirical relation $N_{\rm H}\approx 8.9\times10^{21}E(B-V)$ 
cm$^{-2}$ \citep{foight2016}\footnote{There are only few data points for low $N_{\rm H}$/$E(B-V)$,   
which can lead to potentially unaccounted  uncertainties  of this relation.}.

As noted above, the \psr\ X-ray spectrum alone is best described by the absorbed model
consisting of the non-thermal (PL) and thermal components originating from the pulsar magnetosphere 
and the NS surface, respectively.
To describe the latter, we used the NS magnetic hydrogen atmosphere models with the dipole magnetic field, {\sc nsmdip}, 
which were calculated for NSs with low effective surface temperatures
(see \citealt{XMMPSRJ1957} for detailed model description). 
The models were constructed for several     
NS masses, radii and magnetic field strengths.  
Free parameters are the redshifted effective temperature $T^\infty$, 
the distance $D$ to the pulsar, the angle $\alpha$ between the rotation 
and magnetic axes and the angle $\zeta$ between the rotation axis 
and the line of sight.

To consider where the optical point is located in respect to the X-ray spectral fit alone, we extrapolated the best-fitting models of the X-ray spectra
obtained by \citet{XMMPSRJ1957} to the optical. An example of such
extrapolation for the model {\sc nsmdip1}+PL is presented in 
the top panel of Figure~\ref{fig:psr-models-src}.
The atmosphere model {\sc nsmdip1} was produced for the fixed NS mass of 1.4\msun\ , NS radius of 12.6 km
and magnetic field at the pole of 3$\times$10$^{12}$ G.
One can see that the model strongly overshoots 
the observed optical flux density and thus cannot  describe the multi-wavelength  data.  
For other {\sc nsmdip+PL} models from \citet{XMMPSRJ1957}, which equally well describe the X-ray spectrum,  the situation
is similar. Therefore, below we focus  on the results including 
only {\sc nsmdip1} model.

Then we fitted the X-ray spectra and the optical data point simultaneously. 
We considered the two simplest cases to estimate the properties of the NS thermal emission. 
In case 1, we assume that the pulsar X-ray and optical non-thermal emission can be described by a single PL component.
In case 2, we assume that 
the PL component has a 
negligible contribution
in the optical band.  
Thus, we set the PL normalization to zero 
in the optical band, and 
this allows us to get upper bounds 
on the pulsar thermal emission. 
To estimate fit qualities for  cases 1 and 2, 
we  grouped the X-ray data to ensure 25 counts per energy bin 
and used the  $\chi^2$ fit statistic\footnote{Here we did not
use the \chan\ data due to the very low count statistic.}.  
As a result, we got almost equally acceptable fits with 
reduced $\chi^2$ = 1.28 and  1.15 at 53 degrees of 
freedom (dof) for the cases 1 and 2, respectively.

Since the number of X-ray source counts is not large, in order to get the most robust estimates of 
the model parameters and their uncertainties from spectral fits, 
we regrouped all X-ray spectra to ensure at least 1 count per energy bin and 
 used the $W$-statistics\footnote{The $W$-statistics is the $C$-statistics \citep{1979cash} appropriate for Poisson data with Poisson background; see \url{https://heasarc.gsfc.nasa.gov/xanadu/xspec/manual/XSappendixStatistics.html}}.
For the optical spectral bin, containing $\sim 3500$ source counts, 
the $\chi^2$-statistics 
was applied. 
We then utilised the Bayesian parameter estimation procedure using {\sc pyxspec}
interface and a {\sc python} package {\sc emcee} \citep{emcee2013}. 
The fitting procedure is the same as described in  \citet{XMMPSRJ1957}. 
It includes the relation between the interstellar extinction and 
distance in the \psr\ direction  as a prior, allowing one to get a better estimation on the distance to the pulsar,  
and the constraint $\alpha + \zeta \leq 90$\degs\ 
following from the X-ray pulse profile  analysis.
We derived the best-fitting parameters, which are the maximal-probability density values, 
and credible intervals from the sampled posterior distributions. 
The resulting parameters are presented in Table~\ref{tab:fit} and 
the best-fitting models are shown in the middle and bottom panels of 
Figure~\ref{fig:psr-models-src}. 
Obtained likelihoods are similar to those for the X-ray data fitting 
alone (see Table~\ref{tab:fit}) demonstrating the similar fit qualities.
The angles $\alpha$ and $\zeta$ cannot be constrained 
from the spectral analysis. 
1D and 2D marginal posterior distributions for parameters in case 2
are presented in Figure~\ref{fig:trngl}.

Since we cannot exclude the possibility that the detected optical source 
is a background object, we performed the spectral analysis using the 3$\sigma$ 
detection limit. To include the latter, we created a fictitious 
point with error bars, $0.5f_{3\sigma}\pm 0.5f_{3\sigma}$.
The fit quality and derived parameters are similar to those obtained above since the upper limit 
is comparable to the presumed counterpart flux density.

\citet{XMMPSRJ1957} showed  that  the BB model can also  fit the \psr\ thermal 
spectral component in X-rays implying  the presence   of a hot part of the NS surface 
with the temperature of $\approx 50$ eV.  
Fitting the optical--X-ray spectrum  (binned to ensure
at least 25 counts per energy bin) with the BB+PL model 
results in $\chi^2_\nu$ of 1.35 (dof = 55) that is worse than the respective   fit with the atmosphere model {\sc nsmdip1}+PL (see above). The fit becomes significantly better, $\chi^2_\nu$=1.15 (dof = 53), if we use the   BB+broken PL model.
It suggests a PL spectral break located somewhere in the UV. However, the break  position and the PL slope 
in the optical remain highly uncertain.  
At the same time, the rest of the spectral  parameters obtained in this fit are the same as for the BB+PL 
model applied to the X-ray spectrum alone  \citep{XMMPSRJ1957}. A rough upper bound of the photon index at 
frequencies below the break in this model  is about 1.

We also note, that an additional BB 
component related   with  colder parts of the \psr\ surface may contribute in the optical--UV band of \psr\ but not in X-rays, as observed for other 
NSs, e.g. the `Magnificent Seven' group \citep{RXJ1856.5-3754-2003A&A,kaplan2011}.
However, the parameters of such possible component 
currently cannot be robustly  constrained for \psr\ having only the single optical point.

\begin{table*}
\caption{Results of the optical-X-ray spectral fitting$^\dag$.}
\renewcommand{\arraystretch}{1.2}
\begin{tabular}{llllllllllll}
\hline
Case & \nh,                 & $T^{\infty}$,    & lg$L^\infty$, & $\Gamma$  &  $K$,                  &  $D$, & $-\ln \mathcal{L}$ & $N_{\rm bin}$\\       
     & 10$^{20}$ cm$^{-2}$  & eV               & erg s$^{-1}$  &           &  \phs                  &  pc & \\
\hline 
\multicolumn{8}{c}{X-ray data alone (results from table 1 in \citealt{XMMPSRJ1957})}\\
\hline
     & $3.6^{+0.6}_{-0.5}$ & $21.7^{+2.0}_{-1.7}$   & $30.83^{+0.16}_{-0.14}$ & $1.65^{+0.11}_{-0.10}$ & $4.8^{+0.5}_{-0.3}\times10^{-6}$ & $260^{+109}_{-82}$ & 195.3 & 373\\
\hline
\multicolumn{8}{c}{X-ray data $+$ the optical data point from presumed pulsar}\\
\hline
1    & $3.0^{+0.5}_{-0.7}$ & $23.8^{+2.1}_{-1.5}$ & $30.99^{+0.15}_{-0.11}$ & 1.43$^{+0.04}_{-0.05}$ &  4.13$^{+0.28}_{-0.19}$ $\times$ 10$^{-6}$ & $404^{+151}_{-99}$ & 198.6 &374 \\
2    & $3.6^{+0.5}_{-0.4}$ &$21.1^{+1.3}_{-1.0}$ & $30.79^{+0.10}_{-0.09}$ & 1.66$^{+0.10}_{-0.10}$ &  4.94$^{+0.37}_{-0.34}$ $\times$ 10$^{-6}$ & $215^{+52}_{-22}$ & 195.5 & 374\\
\hline
\end{tabular}
\begin{tablenotes}
\item $^\dag$ 
$L^\infty$ is 
the bolometric thermal luminosity as measured by a distant observer, 
$\Gamma$ is the photon index,  $K$ is the PL normalization, $\ln \mathcal{L}$ is the 
log-likelihood and $N_{\rm bin}$ is the number of spectral bins. 
All errors are at 68 per cent credible intervals. See text for descriptions of cases 1 and 2. 
\end{tablenotes}
\label{tab:fit}
\end{table*}


\section{Discussion} 
\label{sec:discussion}

A faint source was detected within the 1$\sigma$ positional error radius of J1957
with $g'=27.63\pm 0.26$  at the 4$\sigma$ significance. We consider
it as a likely pulsar counterpart. 
Since the source can be a background object,
we also calculated
 the optical upper limit of 27.94 on the pulsar brightness in the 
 $g'$ band.
We performed spectral fits  using 
these values together
with the pulsar X-ray data. 

The \psr\ X-ray spectrum alone can be described by the  
NS atmosphere model with the addition of a non-thermal PL component 
dominating at high energies 
\citep{XMMPSRJ1957}. 
As seen from the upper panel of Figure~\ref{fig:psr-models-src}, simple extrapolation of the X-ray spectral fit possibly implies  the existence 
of a spectral break of the 
PL component between
X-rays and the optical. 

To check this, we first performed the multi-wavelength spectral fit without the break
and obtained an acceptable result (see the middle panel of Figure~\ref{fig:psr-models-src}).
The addition of the optical point to the X-ray data leads to a marginal change of the NS surface
temperature, a flatter PL slope and about two times larger distance (see Table~\ref{tab:fit}). 
The optical emission is mainly dominated by the PL component in this case, the contribution of the thermal component is  $\sim$20 per cent. We estimated the non-thermal
luminosity of \psr\ in the $g'$ band, $\approx$ 1.6 $\times$ 10$^{27}$
\ergs, resulting in  
the  efficiency of transformation of its spin-down energy losses
($\dot{E}=5.3\times10^{33}$ \ergs)
to the optical photons  
of $10^{-6.5}$. These values are comparable with the respective estimates 
for other $\sim$1 Myr old pulsars \citep{2013MNRAS.435.2227Z}.

However, non-thermal spectral components of pulsars typically 
become significantly flatter in the optical than in X-rays  \citep[e.g.][]{Kirichenko2014,PSRJ1741-2054,PSRJ0108-1431,Zharikov2021}.
It is impossible to obtain the spectral slope of J1957 in the optical having only 
a single data point. 
Therefore, we considered an extreme case where the non-thermal component has  a sufficiently strong spectral break  making it  negligible  in the optical.
If so, the optical emission mainly originates from the surface of the cooling NS, which appears to be  reasonable for
 a
cold and nearby pulsar. This is supported by the fact that the  
extrapolation of its  thermal component from X-rays to the optical
is consistent with the optical flux density (see the upper panel of Figure~\ref{fig:psr-models-src}).
As a result (the bottom panel of Figure~\ref{fig:psr-models-src} and the last line in Table~\ref{tab:fit}), the parameters of 
the non-thermal component remain almost the same as for the 
best-fitting model of the X-ray data alone, but the distance has smaller uncertainties. 

The two considered cases shown in the middle and bottom panels of Figure~\ref{fig:psr-models-src}    resulted in  close fit statistics (Table~\ref{tab:fit}), thus we  
cannot exclude or confirm the PL spectral break 
between the optical and X-rays. However, the existence of a spectral break
is more preferable due to the higher likelihood $\mathcal{L}$ (lower $- \ln \mathcal{L}$) in case 2.
We argue 
that 
for all reasonable 
expected breaks   the pulsar parameters presented in Table \ref{tab:fit} should be 
in the ranges between the values provided by these two limiting cases.  
If
the detected source is an  unrelated  background   object, we obtained similar fit results 
for the optical counterpart upper limit, as it 
is comparable 
to the  flux density of the likely counterpart.

Finally, \citet{XMMPSRJ1957} also used the BB model to fit the \psr\ thermal 
spectral component in X-rays, which may be  reminiscent of the emission from 
the solid state star  surface. The results imply the emission from a hot part of the NS surface with the temperature $\approx 50$ eV. 
In this case, the broken PL 
is certainly required to describe the 
multi-wavelength spectrum  suggesting  
 the break somewhere in the UV range.

Our multi-wavelength spectral analysis favours the pulsar nature of the detected optical source, since it provides physically adequate  parameters of \psr\ and its emission. 
To  confirm this,   deep UV observations would be particularly important.  
Based on the middle and bottom panels of Figure~\ref{fig:psr-models-src}, the pulsar flux density is expected to be higher in the UV than in the $g'$ band 
due to the increase of the contribution of the pulsar thermal emission component. 
Even if the contribution of the nonthermal emission component is significant, the counterpart optical-UV  colour should be considerably different from a stellar one. 
Additional data point(s) will 
allow one to select the best spectral model among the considered ones and 
thus to get new convincing constraints on the pulsar parameters.



\section*{Acknowledgements}
 
We are grateful to anonymous referee for useful comments. The work is based on observations made with the Gran Telescopio Canarias, 
installed at the Spanish Observatorio del Roque de los Muchachos 
of the Instituto de Astrof\'isica de Canarias, in the island of La Palma. 
{\sc iraf} is distributed by the National Optical Astronomy Observatory, 
whinomy (AURA) under a cooperative agreement with the National Science Foundation. 
This work has made use of data from the European Space Agency (ESA) mission 
{\it Gaia} (\url{https://www.cosmos.esa.int/gaia}), processed by the {\it Gaia} 
Data Processing and Analysis Consortium 
(DPAC; \url{https://www.cosmos.esa.int/web/gaia/dpac/consortium}).  
Funding for the DPAC has been provided by national institutions, 
in particular the institutions participating in the {\it Gaia} Multilateral Agreement.
DAZ thanks Pirinem School of Theoretical Physics for hospitality.
The work of DAZ, AVK, YAS and AYP was partially  supported by the Russian Foundation for Basic Research,
project 19-52-12013 NNIO\_a. SVZ acknowledges PAPIIT grant IN102120.
The work of VFS was supported by  Deutsche  Forschungsgemeinschaft  (DFG)  (grant WE 1312/53-1) and
was partially supported by the subsidy (project no. 0671-2020-0052) allocated to the Kazan Federal
University for the State assignment in the sphere of scientific activities.

\section*{Data Availability}

The X-ray and optical data are available through their respective data 
archives: \url{https://www.cosmos.esa.int/web/xmm-newton/xsa} for 
\xmm\ data, \url{https://cxc.harvard.edu/cda/} for \chan\ data and \url{https://gtc.sdc.cab.inta-csic.es/gtc/} for GTC.



\bibliographystyle{mnras}
\bibliography{1957arxiv} 




\bsp	
\label{lastpage}
\end{document}